\documentstyle[11pt]{article}  
\newcommand {\bl}{\begin{list}{}{\leftmargin 2em}}
\newcommand {\ite}{\item{}\hspace{-2em}}
\newcommand {\el}{\end{list}}
\newcommand {\beq}{\begin{equation}}
\newcommand {\eeq}{\end{equation}}

\newcommand{\bm}[1]{\mbox{\boldmath $#1$}}
\begin{document}

\hspace*{-2mm}{\footnotesize Presented at the IUGG95 Assembly, Boulder,
Colorado, July 2-14, 1995.}

\begin{center}
\section*{\bf Mixed integer programming for the resolution of
 GPS carrier phase ambiguities }

\vspace{5mm}

 Peiliang Xu \hspace{1mm} Elizabeth Cannon \hspace{1mm} Gerard Lachapelle
\\ Department of Geomatics Engineering, The University of Calgary\\
2500 University Dr. NW, Calgary AB T2N 1N4, Canada

\end{center}

\vspace{2mm}

\hspace*{17mm}\begin{minipage}{13cm}{ {\bf Abstract} \\ {\sl This arXiv upload is to clarify that
the now well-known sorted QR MIMO decoder was first presented in the 1995 IUGG General Assembly. We
clearly go much further in the sense that we directly incorporated reduction into this one step,
non-exact suboptimal integer solution. Except for these first few lines up to this point, this
paper is an unaltered version of the paper presented at the IUGG1995 Assembly in Boulder.

The Ambiguity resolution of GPS carrier phase observables is crucial in high precision geodetic
positioning and navigation applications. It consists of two aspects: estimating the integer
ambiguities in the mixed integer observation
 model and examining whether they are sufficiently accurate to be fixed as
known nonrandom integers. We shall discuss the first point in this paper from
the point of view of integer programming. A one-step nonexact approach is
proposed by employing minimum diagonal pivoting Gaussian decompositions, which
may be thought of as an improvement of the simple rounding-off method, since
the weights and correlations of the floating-estimated ambiguities are
  fully taken  into
account. The second approach is to reformulate the mixed integer least squares
problem into the standard {\bf 0-1} linear integer programming model, which can
then be solved by using, for instance, the practically robust and efficient
simplex  algorithm for linear integer programming. It is
exact, if proper bounds for the ambiguities are given. Theoretical results on
decorrelation by unimodular transformation are given in the form of a theorem.
} }
\end{minipage}

\parindent 4mm

\section{Introduction}
Three types of observables may be derived from tracking GPS satellites:
pseudorange
(code) measurements, raw Doppler shifts (or equivalently range rates) and
carrier phases. They are used at  different levels of accuracy for different
purposes of applications (see e.g Wells et al. 1986; Leick 1990;
Hofmann-Wellenhof et al. 1992; Seeber 1993; Melbourne 1985). The carrier phase
measurements,  together with the
accurate code observables (if available), have been dominating in high
precision geodetic positioning and navigation applications. The mathematical
model can  symbolically be written below
 $$
  {\bf R}  =  {\bf f}_R({\bf X}) + {\bf B}_R\bm{\lambda} + \varepsilon_R
\eqno(1a)  $$
$$  {\bf \Phi}  =  {\bf f}_{\Phi}({\bf X}) + {\bf B}_{\Phi}
\bm{\lambda} + {\bf B}_Z {\bf Z}  + \varepsilon_{\Phi}.
\eqno(1b) $$
Here ${\bf R}$ and ${\bf \Phi}$ are respectively the observables of
pseudoranges and
carrier phases, $\varepsilon_R $ and $\varepsilon_{\Phi}$ are the random errors
of the observables, ${\bf X}$ is the coordinate vector to be estimated, and
${\bf f}_R(.)$
and ${\bf f}_{\Phi}(.)$ are nonlinear functionals of ${\bf X}$. ${\bf B}_R$,
${\bf B}_{\Phi}$ and ${\bf B}_Z$ are the coefficient matrices. $\bm{\lambda}$
is the vector of nuisance parameters such as the  synchronization errors of
receiver and satellite
clocks and ionospheric corrections. If  overparametrization occurs to
$\bm{\lambda}$,   it is generally not estimable (Wells et al. 1987). Thus we
shall  assume that proper reparametrization has been made by, for instance,
choosing proper datum parameters (Wells et al. 1987) or using differencing and
nuisance parameter elimination techniques (see e.g. Goad 1985; Schaffrin \&
Grafarend 1986), to ensure that the remaining
nuisance parameters are estimable. ${\bf Z}$ is the vector of integral
ambiguities
inhered in the carrier phase observables.

Accurate and reliable resolution of the integral ambiguity vector has been
playing a crucial
role in high precision positioning. There are currently many approximate
proposals available  to resolve ${\bf Z}$.
 They may be treated in two categories: simple (sequential) rounding-off of a
real
number to its nearest integer with and/or without using constraint criteria
(Blewitt 1989; Talbot 1991; Hwang 1991; Seeber 1993; Hofmann-Wellenhof et al.
1992), and  searching methods by employing the information on the prior
statistics and geometry (nonlinear functionals and design matrices) of the
observables (Counselman et al. 1981; Remondi 1990, 1991; Frei \& Beutler 1990;
Mader 1990;  Mervart et al. 1994). Betti, Crespi \& Sans\`{o} (1993) recently
proposed a Bayesian approach to resolution of ambiguity.  Chen \& Lachapelle
(1994) proposed a fast ambiguity search filtering approach to reducing the
number of possible candidates in the searching area. It may be worth
noting that the fast rapid
ambiguity  resolution method proposed by Frei \& Beutler seems to have enjoyed
its wide approval.
 A key element of the method is the use of some formal statistics to pick up a
solution. It may be proved that the statistic used for selecting the candidates
of ambiguities is not mathematically
rigorous, since the ambiguity-free and ambiguity-fixed solution vectors are
both derived by using the same set of carrier phase observations. The method
seems quite successful in practice, however.

Recent progress in resolving the integral ambiguity vector has been made by
Teunissen (1994).  His approach consists of three steps: (1) decorrelation of
the floating-estimated  ambiguities by Gaussian transformation, which may be
said to   characterize the novelty of the new approach, (2) searching for
the solution to the transformed integer least squares problem within a
superellipsoid corresponding to a certain level of confidence, and (3)
back-substituting the solution just derived for the ambiguity vector in the
original model. The success of the approach will depend, to a great extent, on
the first two steps. Testing results of the approach can be found in Teunissen
(1994) and de Jonge \& Tiberius (1994).  Decorrelation techniques may be also
 well suited to explain an important finding by Melbourne (1985),   that the
widelane ambiguity is easier to solve, based on the {\em one epoch}
dual frequency carrier phase and code-derived pseudorange model.

The purpose of this paper is to further study the GPS ambiguity resolution
 as a mixed  integer least squares (LS) mathematical programming problem.
Unimodular integer transformation is used to statistically decorrelate the
 floating-estimated  ambiguities, which summarizes the first two conditions of
transformation proposed by Teunissen (1994).  Two methods for
solving the transformed integer LS problem are then proposed. The first one is
to decompose the
transformed positive definite matrix into a lower and an upper triangle by
 choosing the minimum diagonal  elements. In this way, we are sure that a
wrongly selected ambiguity will be penalized. No iterations
are required, thus it should improve  the sum of square of
the residuals  derived by rounding the transformed real
values to their nearest integers.  The second one is
to reformulate the transformed integer LS problem to a quadratic {\bf 0-1}
nonlinear programming, and then further to a  {\bf 0-1} linear integer
programming.   Thus  simplex algorithms can be employed to efficiently solve
the linear integer programming problem, with which one need not test every
  point in the feasible solution  set.

\section{Integer and mixed integer least squares models}
In the application of the GPS system to high precision positioning and
navigation, the GPS satellites have been treated as space targets with known
positions, unless the determination of the satellite orbits is of interest. In
this paper, we assume that the coordinates of the satellites are given, which
can be computed, for instance, from the (precision) ephemerides. Furthermore,
given a set of approximate coordinates of the stations, we can linearize the
observation equations (1a) and (1b) as
 $$
  {\bf y}_R  =  {\bf A}_R \Delta {\bf X} + {\bf B}_R \Delta \bm{\lambda} +
\varepsilon_R   \eqno(2a)  $$
$$  {\bf y}_{\Phi}  =  {\bf A}_{\Phi} \Delta {\bf X}
+{\bf B}_{\Phi} \Delta \bm{\lambda} + {\bf B}_Z \Delta {\bf Z} +
\varepsilon_{\Phi} \eqno(2b) $$
where  $${\bf y}_R = {\bf R} - {\bf f}_R({\bf X}_0) - {\bf B}_R \bm{\lambda}_0
\eqno(3a) $$
  $${\bf y}_{\Phi} = {\bf \Phi} - {\bf f}_{\Phi}({\bf X}_0)  - {\bf B}_{\Phi}
\bm{\lambda}_0 - {\bf B}_Z {\bf Z}_0 \eqno(3b) $$
$$ \Delta {\bf X} = {\bf X} - {\bf X}_0; \hspace{2mm} \Delta \bm{\lambda} =
\bm{\lambda} - \bm{\lambda}_0
\eqno(3c) $$
$$ \Delta {\bf Z} = {\bf Z} - {\bf Z}_0. \eqno(3d) $$
${\bf X}_0$ and $\bm{\lambda}_0$ are the approximate values of ${\bf X}$ and
$\bm{\lambda}$,
respectively. ${\bf Z}_0$  are integer approximate values of ${\bf Z}$, and
thus $\Delta {\bf Z}$
 remain integral.

Rewriting the linearized observation equations (2a) and (2b) in matrix form,
together with the statistical information on the observables, we have
$$ \left[ \begin{array}{l} {\bf y}_R \\ {\bf y}_{\Phi} \end{array} \right] =
\left[ \begin{array}{l} {\bf A}_R \\ {\bf A}_{\Phi} \end{array} \right] \Delta
{\bf X} +
\left[ \begin{array}{l} {\bf B}_R \\ {\bf B}_{\Phi} \end{array} \right] \Delta
\bm{\lambda} +
\left[ \begin{array}{l} {\bf 0} \\ {\bf B}_Z \end{array} \right] \Delta
{\bf Z} +
\left[ \begin{array}{l} \varepsilon_R \\ \varepsilon_{\Phi} \end{array} \right]
  \eqno(4a) $$
$$ D\left[ \begin{array}{l} {\bf y}_R \\ {\bf y}_{\Phi} \end{array} \right] =
 \left[ \begin{array}{ll} {\bf P}_R & {\bf 0} \\ {\bf 0} & {\bf P}_{\Phi}
\end{array} \right]^{-1} \sigma^2. \eqno(4b) $$
Here ${\bf P}_R$ and ${\bf P}_{\Phi}$ are respectively the weight matrices of
the observables ${\bf y}_R$ and ${\bf y}_{\Phi}$, $\sigma^2$ is the scalar
variance component.

Since the main interest of this paper is to discuss the mixed integer LS
problem, we do not need to discriminate between the position unknowns ${\bf X}$
and the nuisance parameters $\bm{\lambda}$. Without loss of generality,
therefore, we can
simplify the model (4) as the following standard mixed real-integer (or simply
integer in the rest of the paper) observation equations,
$$ {\bf y} = {\bf A} \bm{\beta} + {\bf B} {\bf z} + \varepsilon \eqno(5a) $$
$$ D({\bf y}) = {\bf P}^{-1} \sigma^2 \eqno(5b) $$
where
$$ {\bf y} = \left[ \begin{array}{l} {\bf y}_R \\ {\bf y}_{\Phi} \end{array}
\right];
\hspace{2mm} \varepsilon = \left[ \begin{array}{l} \varepsilon_R \\
\varepsilon_{\Phi} \end{array} \right]  $$
$$ {\bf A} = \left[ \begin{array}{ll} {\bf A}_R & {\bf B}_R \\
    {\bf A}_{\Phi} & {\bf B}_{\Phi} \end{array} \right]; \hspace{1mm}  {\bf
\beta} = \left[
\begin{array}{l} \Delta {\bf X} \\ \Delta \bm{\lambda} \end{array} \right] $$
$$ {\bf B} = \left[ \begin{array}{l} {\bf 0} \\ {\bf B}_Z \end{array}
\right];
\hspace{2mm}
 {\bf z} = \Delta {\bf Z}  $$
$$ {\bf P} =  \left[ \begin{array}{ll} {\bf P}_R & {\bf 0} \\ {\bf 0} & {\bf
P}_{\Phi} \end{array} \right]. $$
The matrices ${\bf A}$ and ${\bf B}$ are full of column rank, respectively.

Applying the least squares criterion to (5), we have
\setcounter{equation}{5}
\beq
min: \hspace{3mm} F = ( {\bf y} - {\bf A} \bm{\beta} - {\bf B} {\bf z} )^T
{\bf P} ( {\bf y} - {\bf A} \bm{\beta} - {\bf B} {\bf z} ),
\eeq
which is   the mixed integer LS problem. (6) was also called the constrained LS
 problem by Teunissen (1994).  Since the variables ${\bf z}$ are discrete,
we cannot use the conventional method by differentiating the objective function
$F$ with respect to the variables $\bm{\beta}$ and ${\bf z}$ in order to form
the normal
equation and then solve for them. Instead, however, we differentiate $F$ with
respect to $\bm{\beta}$ and let it equal zero, leading to
$$ \frac{\partial F}{\partial \bm{\beta} } = -2 {\bf A}^T{\bf P} ({\bf
y} - {\bf A} \bm{\beta} - {\bf B} {\bf z} ) = {\bf 0} $$
or
$$ {\bf A}^T{\bf P}{\bf A} \bm{\beta} = {\bf A}^T{\bf P}({\bf y}- {\bf
B}{\bf z}). $$
Hence
\beq
\bm{\beta} = ({\bf A}^T{\bf P}{\bf A})^{-1} ({\bf y} - {\bf B}{\bf z}).
\eeq

Substituting (7) into (5) and rearranging it yield
$$ {\bf y}_1 = {\bf Q}{\bf P} {\bf B} {\bf z} + \varepsilon_1  \eqno(8a) $$
$$ D[{\bf y}_1] = [{\bf P}^{-1} - {\bf A}({\bf A}^T{\bf P}{\bf A})^{-1} {\bf
A}^T] \sigma^2 = {\bf Q} \sigma^2 \eqno(8b) $$
where $$ {\bf y}_1 = [{\bf I}-{\bf A}({\bf A}^T{\bf P}{\bf A})^{-1} {\bf
A}^T{\bf P}] {\bf y} = {\bf Q} {\bf P} {\bf y} $$
 $$ {\bf Q}={\bf P}^{-1} - {\bf A}({\bf A}^T{\bf P}{\bf A})^{-1} {\bf A}^T. $$

Applying the LS method to (8), we have
\setcounter{equation}{8}
\begin{eqnarray}
min: \hspace{2mm} F_1 &  = & ( {\bf y}_1 - {\bf Q}{\bf P}{\bf B} {\bf z})^T
{\bf Q}^- ( {\bf y}_1 - {\bf Q}{\bf P}{\bf B} {\bf z}) \nonumber \\
          &  = & ({\bf y} - {\bf B} {\bf z})^T{\bf P}{\bf Q} {\bf Q}^- {\bf
Q}{\bf P} ({\bf y} - {\bf B}{\bf z}) \nonumber \\
      &  = & ({\bf y} - {\bf B} {\bf z})^T  {\bf P} {\bf Q} {\bf P}   ({\bf y}
- {\bf B} {\bf z}) \nonumber \\
      & =  & {\bf y}^T{\bf P}{\bf Q}{\bf P}{\bf y} - 2 {\bf y}^T{\bf P}{\bf
Q}{\bf P}{\bf B} {\bf z} + {\bf z}^T{\bf B}^T{\bf P}{\bf Q}{\bf P}{\bf B}{\bf
z}.
\end{eqnarray}

The objective function $F_1$ can further be rewritten as
\beq
 min: \hspace{2mm} F_1 = ( {\bf z} - \hat{{\bf z}})^T {\bf H} ({\bf z} -
\hat{{\bf z}}) + {\bf y}^T{\bf P}{\bf Q}[{\bf Q}^- -
 {\bf P}{\bf B}{\bf H}^{-1}{\bf B}^T{\bf P}] {\bf Q}{\bf P}{\bf y}
\eeq
where $$ \hat{{\bf z}} = {\bf H}^{-1}{\bf B}^T{\bf P}{\bf Q}{\bf P}{\bf y} $$
 $$ {\bf H} = ({\bf B}^T{\bf P}{\bf Q}{\bf P}{\bf B}). $$
Here $\hat{{\bf z}}$ can readily be proved to be the floating LS estimate of
the
ambiguity vector $\Delta {\bf Z}$ with covariance matrix ${\bf
H}^{-1}\sigma^2$. Since $ {\bf y}^T{\bf P}{\bf Q}[{\bf Q}^- -  {\bf P}{\bf
B}{\bf H}^{-1}{\bf B}^T{\bf P}] {\bf Q}{\bf P}{\bf y}
$ is constant, the objective function (10) is
equivalent to (Teunissen 1994; de Jonge \& Tiberius 1994)
\beq
min: \hspace{3mm} F_2 = ( {\bf z} - \hat{{\bf z}})^T {\bf H} ({\bf z} -
\hat{{\bf z}}),
\eeq
which is the standard integer LS problem.

It is now clear that the solution to the original mixed integer LS problem (6)
depends solely on that of the standard integer LS problem (11). Denote
the integer solution of ${\bf z}$ to (11) by $\hat{{\bf z}}^{IN}$. Substituting
it into
(7), we can then obtain the LS estimates of the real parameters $\bm{\beta}$
without
much effort.

\section{Unimodular transformation}
In resolution of GPS carrier phase ambiguities, one of the most difficult
points is to handle strong correlation of the matrix ${\bf H}$. Searching for
an    acceptable (and/or hopefully optimal) solution of {\bf z} is arduous, if
it is solely based on the strong correlation matrix ${\bf H}$, since testing a
large number of combinations would have to be done.  Roughly speaking, the
total number of combinations required
  is computed by $\prod n_i$, where $n_i$ is the number of integer points on
 an interval of line for the {\em i}th ambiguity, centred at the $\hat{ z}_i$
and corresponding to a significance level (see e.g Frei \& Beutler 1990).
However, if the matrix ${\bf H}$ is
diagonal, one can simply round the floating values $\hat{{\bf z}}$ off to the
nearest
integers, which are the integer solution of ${\bf z}$. Therefore, an idea
would emerge naturally, that one works with a decorrelated weight matrix
instead of ${\bf H}$.

Such a technique was proposed recently by Teunissen (1994) (see also de Jonge
\& Tiberius 1994). His basic idea is to transform the ``observables''
$\hat{{\bf z}}$
 by ${\bf G}$ into the new ones $\hat{\bf z}_1 =( {\bf G}^T\hat{{\bf
z}})$, and then work with the LS integer  problem
\beq
min: \hspace{3mm} F_3 = ( {\bf z}_1 - \hat{\bf z}_1)^T{\bf H}_1 ( {\bf z}_1 -
\hat{\bf z}_1).
\eeq
Here $ {\bf H} = {\bf G}{\bf H}_1{\bf G}^T$. The transformation matrix ${\bf
G}$ has to satisfy the following
three conditions: (1) integer elements; (2) volume preservation; and (3)
decorrelation of ${\bf H}$ into ${\bf H}_1$. More details can be found in
Teunissen (1994).

Before proceeding, we shall define the unimodular matrix (see e.g. Nemhauser \&
Wolsey 1988).

{\bf Definition 1.} A square matrix ${\bf G}$ is said to be {\em unimodular} if
it is integral
and if the absolute value of its determinant is equal to unity, {\em i.e.}
$|det({\bf G})|=1$.

The inverse of a unimodular matrix is also unimodular, since
$|det({\bf G}^{-1})| = 1/|det({\bf G})| = 1$, and because
$$ {\bf G}^{-1} = \bar{{\bf G}}/det({\bf G}) = \pm \bar{{\bf G}}.$$
Here $\bar{{\bf G}}$ is the {\em adjoint matrix} of ${\bf G}$, whose elements
are derived  only by using the operations of integer multiplication,
substraction and addition,
and thus integer. The sign before $\bar{{\bf G}}$ depends on the determinant of
 ${\bf G}$. The second property of unimodular matrices is that the
product of two unimodular matrices is unimodular.  It is also clear that any
unimodular transformation of
an integer vector is an integer vector, too.

By employing the concept of the unimodular matrix, we can summarize the first
two conditions suggested by Teunissen (1994) by stating that the transformation
 ${\bf G}$ is unimodular. It should be noted that there was a misunderstanding
of Teunissen's second condition  of volume preservation. Volume preservation
does not imply  the  preservation of the number of grid points.
 A simple  example is that a unit circle centred at the origin has five
grid points, while an ellipse of the same center with major axis 1.5 and minor
axis 2/3 encloses only three grid points.

Integer Gaussian decomposition was employed by Teunissen (1994), that indeed
decorrelates the matrix ${\bf H}$. What now seems to be done is to
mathematically prove that we can always decorrelate the matrix ${\bf H}$ by
using a finite number of  unimodular transformations to the extent that the
correlation coefficient of any two random variables is always less than or
equal to 1/2. In order to do so, we need the following lemma on the
inequality  of matrix determinant.

{\bf Lemma 1:} For any positive definite matrix ${\bf A}$, the following
inequality
\beq
det({\bf A}) \leq \prod a_{ii}
\eeq
holds true. Here $a_{ii}$ are the diagonal elements of ${\bf A}$.

\hspace*{-6mm} \bm{Proof.} A positive definite matrix ${\bf A}$ can be written
by
Choleski decomposition as
$$ {\bf A} = {\bf L}{\bf L}^T $$
where $ l_{ii} = (a_{ii} - \sum \limits^{i-1} \limits_{j=1} l_{ij}^2 )^{1/2} >
0.$  Thus we have
\begin{eqnarray*}
 det({\bf A}) & = & \prod l_{ii}^2 \nonumber \\
        & = & \prod (a_{ii} - \sum \limits^{i-1} \limits_{j=1} l_{ij}^2 )
\nonumber \\
       & \leq & \prod a_{ii},
\end{eqnarray*}
since $\sum \limits^{i-1} \limits_{j=1} l_{ij}^2 \geq 0$. $\Box$

{\bf Theorem 1:} For any positive definite matrix ${\bf A}$, there exists a
unimodular
 matrix ${\bf G}$ such that
\beq
{\bf A} = {\bf G}{\bf H}{\bf G}^T.
\eeq
Here ${\bf H}$ is positive definite, too, and satisfies
\beq
|h_{ij}| \leq \frac{1}{2} \  min(h_{ii}, \ h_{jj}) \hspace*{5mm} \forall\ i, j
\hspace{2mm} \& \hspace{3mm} i \neq j.
\eeq

\hspace*{-6mm} \bm{Proof.} Suppose, without loss of generality, that for any
three elements $a_{ii}$, $a_{jj}$ and $a_{ij}$ of the positive definite matrix
${\bf A}$, we have $|a_{ij}|/min(a_{ii}, a_{jj}) > 1/2$. Then construct the
unimodular matrix
$$ {\bf G}_1 = \left[ \begin{array}{lclclcl}
   1 &   &   &    &    &    &    \\
     & \ddots &   &   &   &   &   \\
     &   &  1 &   &    &    &    \\
     &   & \vdots & \ddots &    &    &   \\
     &   & -[a_{ij}/a_{ii}]_{in} & \cdots &  1  &    &    \\
     &   &    &    &    &   \ddots &    \\
     &   &    &    &    &      &   1
\end{array}  \right] \eqno(16a) $$
if $ a_{ii} \leq a_{jj}$, or
$$ {\bf G}_1 = \left[ \begin{array}{lclclcl}
   1 &   &   &    &    &    &    \\
     & \ddots &   &   &   &   &   \\
     &   &  1 & \cdots  &  -[a_{ij}/a_{jj}]_{in}  &    &    \\
     &   &  & \ddots & \vdots   &    &   \\
     &   &  &  &  1  &    &    \\
     &   &    &    &    &   \ddots &    \\
     &   &    &    &    &      &   1
\end{array} \right] \eqno(16b) $$
if $ a_{jj} < a_{ii}$. Here $[\  x \ ]_{in}$ is the operation to round the
floating number $x$ to its nearest integer.

Upon left- and right-multiplying ${\bf A}$ by  the unimodular matrix ${\bf
G}_1$ and its transpose respectively, the larger diagonal element is then
reduced to
\setcounter{equation}{16}
\beq max(a_{ii}, a_{jj}) - 2[a_{ij}/a_{min}]_{in}a_{ij}  +
a_{min}[a_{ij}/a_{min}]^2_{in}
\eeq where $a_{min} = min(a_{ii},a_{jj}). $  Repeating the same procedure to
any pair of diagonal elements, we have
\beq
{\bf A}_n = {\bf G}_n...{\bf G}_1{\bf A}{\bf G}_1^T...{\bf G}_n^T.
\eeq

Now suppose that we cannot reach the equation (14) and the inequality (15) by
employing a finite number of unimodular matrices of the form (16), then we keep
applying the same procedure to ${\bf A}_n$. By expression (17), it is clearly
true that the minimum diagonal element of the reduced matrix, say ${\bf A}_m$
now, has no lower
bound. It means that the minimum element can be arbitrarily small, which
further implies by Lemma 1 that
\beq
det({\bf A}_m) \leq \prod a^m_{ii} < const,
\eeq
where $a^m_{ii}$ are the diagonal elements of ${\bf A}_m$, $const$ is any
positive constant. Since
unimodular transformation does preserve the determinant, we have $det({\bf
A}_m) =
det({\bf A})$ --- a finite constant,
which clearly contradicts (19). Therefore, we must be able to reach the
condition (15). On the other hand, all the transformation matrices involved are
unimodular, their product is unimodular, too. Denoting the final reduced matrix
 by ${\bf H}$, which satisfies the condition (15), and the product of all the
unimodular matrices by ${\bf G}_t$, we have
\beq
{\bf H} = {\bf G}_t {\bf A} {\bf G}_t^T
\eeq
or
\beq
{\bf A} = {\bf G}{\bf H}{\bf G}^T.
\eeq
Here ${\bf G}(={\bf G}_t^{-1})$ is unimodular.   The proof that the matrix
${\bf H}$ is positive definite is trivial. $\Box$

\section{Two approaches to the integer LS problem}
 The integer LS problem is simply an integer quadratic programming issue. One
can use
 any advanced integer programming algorithm (Parker \& Rardin 1988) to solve
this problem. Essentially, no bounds for the integer unknowns are required and
 no statistical techniques needed to reduce the number of possible candidates.
More on these aspects and proper validation criteria for fixing the carrier
phase ambiguities will be presented in a future paper.

 Though the techniques to be presented below require no decorrelation as an
assumption, and consider that the original and the transformed LS integer
problems are of the same form, the following discussion will be based on the
transformed model, without loss of generality.   After the weight matrix of the
floating-estimated ambiguity vector is
decorrelated, one can either simply round the transformed floating numbers off
to their nearest integers, or employ  searching techniques to find the
``optimal'' solution within a superellipsoid under a certain level of
confidence (Teunissen 1994). In what follows, we shall develop two approaches
to resolve the ambiguities of the transformed integer LS problem.

\subsection{A one-step nonexact approach by minimum diagonal pivoting Gaussian
decomposition  }
Instead of
directly applying the simple rounding-off method to (12), which ignores any
correlation information on the floating-estimated ambiguities, we propose an
alternative one-step approach, based on the weights and correlations of the
transformed ambiguities.  The basic idea is to resolve the integer ambiguities
according to their weights and correlations. As long as some of ambiguities are
resolved, their
correlations with other unfixed floating ambiguities are employed and  the next
 ambiguity corresponding to the large weight is to be determined.

In order to realize the above procedure, we have to decompose the positive
 definite matrix ${\bf H}_1$ carefully. Here we employ Gaussian decomposition by
selecting the minimum diagonal element. The decomposition procedure consists of
the following steps:
\begin{itemize}
\item Selecting the minimum element among all the undecomposed diagonal
elements;
\item Exchanging the rows and the columns;
\item Performing Gaussian decomposition;
\item Replacing the square root of the decomposed element $h_{1(ii)}'$ at the
corresponding position of the factor matrix ${\bf L}$; If the decomposition is
not
completed, then go to the first step. Otherwise, the decomposition is finished.
 \end{itemize}

In mathematical language, we can express the matrix ${\bf H}_1$ as
\beq
{\bf H}_1 = {\bf P}_h{\bf L}{\bf L}^T{\bf P}_h^T
\eeq
where ${\bf P}_h$ is the permutation matrix which represents the exchange of
the rows
and columns during the decomposition. A significant characteristic of this
decomposition is to keep the diagonal elements of the lower triangular matrix
${\bf L}$  in the increasing order as far as possible.

Inserting ${\bf H}_1$ in (22) into (12), we have the objective function
\begin{eqnarray}
min: \hspace{3mm} F_3 & = & ({\bf z}_1 - \hat{\bf z}_1)^T{\bf P}_h{\bf L} {\bf
L}^T{\bf P}_h^T({\bf z}_1 - \hat{\bf z}_1)
\nonumber \\
   & = & ({\bf z}_2 - \hat{\bf z}_2)^T{\bf L}{\bf L}^T({\bf z}_2 -
\hat{\bf z}_2)
\end{eqnarray}
where
\beq
{\bf z}_2 = {\bf P}_h^T{\bf z}_1; \hspace{3mm} \hat{\bf z}_2 = {\bf
P}_h^T\hat{\bf z}_1.
\eeq

Since the factor matrix ${\bf L}$ is lower triangular, we can rewrite (23) as
\beq
min: \hspace{3mm} F_4 = \sum_{i=1}^{t_z} \ [\sum_{j=i}^{t_z}
l_{ji}(z_{2(j)}-\hat{z_{2(j)}})\ ]^2.
\eeq
Here $t_z$ is the dimension of the ambiguity vector ${\bf z}$ (or ${\bf z}_2$).
The solution to the objective function $F_2$ can now be derived by minimizing
\beq
|\sum_{j=i}^{t_z} l_{ji}(z_{2(j)}-\hat{z}_{2(j)})|, \hspace{3mm} \forall\ i.
\eeq

Hence the one-step nonexact integer ambiguity solution is immediate
\beq
\hat{ z}_{2(i)}^{IN} = \left[ \frac{l_{ii} \hat{ z}_{2(i)} - \sum_{j=i+1}^{t_z}
l_{ji}(\hat{z}_{2(j)}^{IN}-\hat{z}_{2(j)}) }{l_{ii}} \right]_{in}
\eeq
for all $i$.

By back substituting the integer solution $\hat{{\bf z}}_2^{IN} =
(\hat{z}_{2(1)}^{IN}, \hat{z}_{2(2)}^{IN}, ..., \hat{z}_{2(t_z)}^{IN})^T$, we
have the final  solution of the integer ambiguities ${\bf z}$, which is denoted
by $\hat{\bf z}^{IN}$,
\beq
\hat{{\bf z}}^{IN} = {\bf G}^{-T}{\bf P}_h\ \hat{{\bf z}}_2^{IN}.
\eeq
\subsection{0-1 quadratic integer programming}
An obvious aim of applying the decorrelation technique to the original integer
LS problem is the alleviation of the computational burden for finding the
optimal
ambiguity solution. When it is translated into the case of searching
techniques,  we expect that the total number of candidate grid points to be
tested should be significantly reduced. Suppose that for the transformed
integer LS problem (12) (${\bf H}_1$ satisfies the conditions of Theorem 1), we
have
to search for the optimal integer ambiguity resolution within the hard bounds
\beq
m_i^0 \leq z_{1(i)} \leq m_i^1, \hspace{3mm} \forall \hspace{2mm} i
\eeq
or in another form,
\beq
z_{1(i)} \in [m_{1i}(=m_i^0), \ m_{2i}, \ ..., \ m_{1s_i}(=m_i^1)].
\eeq
Here $z_{1(i)}$ is the {\em i}th integer component of the integer vector ${\bf
z}_1$,
 $m_{1i}$, $m_{2i}$, .., and $m_{1s_i}$ are the contiguous integers --- the
candidate points of $z_{1(i)}$ with the lower integer bound  $m_i^0$ and the
upper integer bound $m_i^1$. Thus our mixed integer LS problem has been
 reduced to a quadratic integer programming problem with simple integer
constraints.

In what follows we shall further reformulate it by a {\bf 0-1} quadratic
integer programming model. It has been shown by Parker \& Rardin (1988)  that
the integer variable $z_{1(i)}$   can be represented with $r_i$ {\bf 0-1}
variables, {\em i.e.}
\beq
z_{1(i)}  =  m_i^0 + \sum_{j=0}^{r_i-1} 2^j\ b_{i(j)}, \hspace{2mm} \forall
\hspace{2mm} i
\eeq
where $b_{i(j)}$ are {\bf 0-1} integer (binary) variables, $r_i = [log_2(m_i^1
- m_i^0)]_s + 1$,  and $[\ .\ ]_s$ stands for the integer not larger than the
positive number in brackets.

Rewriting all the integer variables $z_{1(i)}$ in matrix form, we have
\beq
{\bf  z}_1 = {\bf m}^0 + {\bf A}_1 {\bf b}
\eeq where the matrix ${\bf A}_1$ is integral with elements $2^k$,  $$ {\bf
m}^0 = (m_1^0,\ m_2^0, \ ...,\ m_{t_z}^0)^T $$
 $$ {\bf b} = ( {\bf b}_1^T, \ {\bf b}_2^T, \ ...,\  {\bf b}_{t_z}^T)^T $$
 $${\bf b}_i = (b_{i(0)}, \ b_{i(1)}, \ ...,\ b_{i(r_i-1)})^T. $$
Furtheron, inserting (32) into the objective function (12) yields
\beq
min: \hspace{2mm} F_3 = ( {\bf A}_1 {\bf b} + {\bf m}^0 - \hat{{\bf
z}}_1)^T{\bf H}_1 ( {\bf A}_1 {\bf b} + {\bf m}^0 - \hat{{\bf z}}_1)
\eeq
subject to $b_k = 0$ or $1$ for all $k$.

The objective function (33) is equivalent to
\begin{eqnarray}
min: \hspace{2mm} F_3 & = & ({\bf m}^0 - \hat{{\bf
z}}_1)^T{\bf H}_1 ({\bf m}^0 - \hat{{\bf z}}_1) + 2({\bf m}^0 - \hat{{\bf
z}}_1)^T {\bf H}_1 {\bf A}_1 {\bf b} \nonumber \\
    &   & + {\bf b}^T{\bf A}_1^T {\bf H}_1 {\bf A}_1 {\bf b}.
\end{eqnarray}
\subsection{0-1 linear integer  programming}
In this subsection, we shall further reformulate the {\bf 0-1} quadratic
programming (34) into a {\bf 0-1} linear integer programming problem by using
the linearization technique.  The basic idea of the linearization technique is
to introduce a new variable to replace the nonzero quadratic term $b_ib_j$.
Thus the {\bf 0-1} quadratic programming problem becomes linear. Since the new
variables are obviously binary, all the variables in the linear programming
model to be  reformulated below are binary, too.

Denoting $$ v_k = b_ib_j, \hspace{2mm} k=(i-1)i/2 + j, \hspace{2mm} i\geq j $$
and taking the following relations  $$ b_i^2 = b_i$$into account, we have
 $$ min: \hspace{2mm} F_4  =  ({\bf m}^0 - \hat{{\bf
z}}_1)^T{\bf H}_1 ({\bf m}^0 - \hat{{\bf z}}_1) + \sum_{i=1}^{t_v} c_iv_i
 \eqno(35a) $$ subject to the following constraints,
 $$ v_i = 0  \vee 1   \eqno(35b) $$
  $$ v_k \geq v_{ki} + v_{kj} - 1  \eqno(35c) $$
$$ v_k \leq v_{ki} \eqno(35d) $$
$$ v_k \leq v_{kj} \eqno(35e) $$
$$ ki = i(i+1)/2; \hspace{2mm} kj = j(j+1)/2. $$
Here $t_v$ is the dimension of the {\bf 0-1} integer vector
 $${\bf v}=(v_1, v_2, \ ...,\ v_{t_v})^T. $$

Since the first term in the objective function (35a) is constant, it is
equivalent to
$$ min: \hspace{2mm} F_4 =  \sum_{i=1}^{t_v} c_iv_i  \eqno(36a) $$
 subject to the constraints (35b $\sim$ e).  (36) is
obviously of the standard form of the {\bf 0-1} linear integer programming.
 It can be solved by using any standard algorithms for {\bf 0-1} linear
programming (Pardalos \& Li 1993; Nemhauser \& Wolsey 1988; Parker \& Rardin
1988; The People University of China 1987). However, the algorithm aspects for
the program (36) will not be discussed here.

\section{Concluding remarks}
GPS carrier phase and pseudorange observables are essentially a nonlinear mixed
  integer observation model. If the GPS satellites are treated as space known
targets, the model is regular. Given a set of approximate values of the unknown
 parameters such as the position coordinates and integer ambiguities, the
nonlinear model is linearized. Estimating the parameters in the linearized
mixed integer model is equivalent to solving a mixed integer LS problem (if the
LS principle is employed), which can be further reduced into a standard integer
LS programming.

It has been recognized that one of the difficulties in correctly estimating the
integer ambiguities is due to the correlations of the floating-estimated
ambiguities.
A decorrelation technique has been proposed by Teunissen (1994), based on
Gaussian decomposition. We have further proved mathematically that there exists
a unimodular  matrix such that (14) and (15) hold true, which may be thought of
as a theoretical summary (and extension) of some of the results in Teunissen
(1994).

Two approaches are then proposed to solve the standard linear integer LS
problem (12) from the point of view of integer programming theory. The first
approach is to Gauss-decompose the matrix ${\bf H}_1$
by selecting the minimum diagonal elements. In other words, we are
estimating the integer ambiguities according to the magnitudes of the weights
of the floating-estimated ambiguities and their correlations (as far as
possible). It may be thought to be an improvement of the simple rounding-off
method.  No iterations are required. It should be noted, however, that this
method is one-step nonexact. The extent of approximation should be further
investigated. The second approach is to reformulate the mixed
integer LS problem into a {\bf 0-1} linear integer programming model. Thus any
standard algorithms for linear integer programming problems can be employed.
The method will result in the exact integer solution of the ambiguities to the
original mixed integer problem, if
proper bounds for the integer unknowns in the transformed model (12) are given.
 Testing of the techniques with real data is under way.
\vspace{2mm}

\hspace*{-4mm}{\bf Acknowledgements:} Much research for this paper was
conceived and done, while PX was a  research fellow
of {\em Alexander von Humboldt} foundation at Stuttgart University with Prof.
Dr. Erik Grafarend as his host. The support from the foundation and the very
friendship of his host are most appreciated.

\section*{References}
\bl
\ite Betti B., Crespi M. \& Sans\`{o} F., (1993): A geometric illustration of
ambiguity in GPS theory and a Bayesian approach, Manus. Geod., 18, 317-330
\\[-6mm]
\ite Blewitt G., (1989): Carrier phase ambiguity resolution for the Global
Positioning System applied to geodetic baselines up to 2000 km, J. geophys.
Res., B94, 10187-10203  \\[-6mm]
\ite Chen D.S. \& Lachapelle G., (1994): A comparison of the FASF and least
squares search algorithms for ambiguity resolution on the fly, In: Proc. Symp.
Kinematic Sys. in Geodesy, Geomatics and Navigation, Banff, Canada, Aug. 30 -
Sept. 2, 1994, pp.241-253 \\[-6mm]
\ite Counselman C.C. \& Gourevitch S.A., (1981): Miniature inteferometer
terminals for earth surveying: ambiguity and multipath with the Global
Positioning System, IEEE Trans. Geosc. Rem. Sen., GE-19, 244-252 \\[-6mm]
\ite Goad C., (1985): Precise relative position determination using Global
Positioning System carrier phase measurements in a nondifference mode, Proc.
1st int. symp. on precise positioning with GPS, Rockville, Maryland, pp.347-356
 \\[-6mm]
\ite Frei E. \& Beutler G., (1990): Rapid static positioning based on the fast
ambiguity resolution approach ``FARA'': theory and first results, Manus. Geod.,
15, 325-356 \\[-6mm]
\ite Hofmann-Wellenhof B., Lichtenegger H. \& Collins J., (1992): GPS ---
theory and practice, Springer-Verlag, Wien \\[-6mm]
\ite Hwang P.Y.C., (1991): Kinematic GPS for differential positioning:
resolving integer ambiguities on the
fly, NAVIGATION (J. Inst. Navigation), 38, No.1 \\[-6mm]
\ite de Jonge P.J. \& Tiberius C., (1994): A new GPS ambiguity estimation
method based on integer least squares, preprint for int. symp. on differential
satellite navigation systems (DSNS94), London \\[-6mm]
\ite Leick A., (1990): GPS satellite surveying, John Wiley \& Sons, New York
\\[-6mm]
\ite Mader G.L., (1990): Ambiguity function techniques for GPS phase
initialization and kinematic solutions,  Proc. 2nd int. symp. on
precise positioning with GPS, Ottawa, Canada, Sept., pp.1234-1247 \\[-6mm]
\ite Melbourne W., (1985): The case for ranging in GPS-based geodetic systems,
Proc. 1st int. symp. on precise positioning with GPS, Rockville, Maryland,
 April 15-19, pp.373-386  \\[-6mm]
\ite Mervart L., Beutler G., Rothacher M. \& Wild U., (1994): Ambiguity
resolution strategies using the results of the International GPS Geodynamics
Service (IGS), Bull. G\'{e}od., 68, 29-38 \\[-6mm]
\ite Nemhauser G. \& Wolsey L., (1988): Integer and combinatorial optimization,
 John Wiley \& Sons, New York \\[-6mm]
\ite Pardalos P. \& Li Y., (1993): Integer programming, in: Computational
Statistics (ed. C.R. Rao), North Holland, Amsterdam, pp.279-302 \\[-6mm]
\ite Parker R. \& Rardin R., (1988): Discrete optimization, Academic Press, New
York \\[-6mm]
\ite The People University of China (PUC), (1987): Introduction to operations
research, The PUC Press, Beijing (in Chinese) \\[-6mm]
\ite Remondi B.W., (1990): Pseudo-kinematic GPS results using the ambiguity
function method, NOAA Tech. Memo. NOS NGS 52, Rockville, MD, also in:
NAVIGATION (J. Inst. Navigation), 38, No.1 \\[-6mm]
\ite Remondi B.W., (1991): Kinematic GPS results without static initialization,
 NOAA Tech. Memo. NOS NGS 55, Rockville, MD \\[-6mm]
\ite Schaffrin B. \& Grafarend E., (1986): Generating classes of equivalent
linear models by nuisance parameter elimination, Manus. Geod., 11, 262-271
\\[-6mm]
\ite Seeber G., (1993): Satellite geodesy, Walter de Gruyter, Berlin \\[-6mm]
\ite Talbot N., (1991): High-precision real-time GPS positioning concepts:
modeling and results,  NAVIGATION (J. Inst. Navigation), 38, No.2 \\[-6mm]
\ite Teunissen P., (1994): A new method for fast carrier phase ambiguity
estimation, Proc. IEEE PLANS94, Las Vegas, Nevada, April 11-15, pp.562-573
\\[-6mm]
\ite Wells D., Beck N., Delikaraoglou D., Kleusberg A., Krakiwsky E.,
Lachapelle G.,
Langley R., Nakiboglu M., Schwarz K.P., Tranquilla J.M. \& Vanicek P., (1986):
Guide to GPS positioning, Canadian GPS Associates, Fredericton, Canada \\[-6mm]
\ite Wells D., Lindlohr W., Schaffrin B. \& Grafarend E., (1987): GPS design:
undifferenced carrier beat phase observations and the fundamental differencing
theorem, UNB Tech. Report No.116, Fredericton, Canada \\[-6mm]
\el

\end{document}